\documentclass[11pt]{article}
\usepackage[utf8]{inputenc}
\usepackage[a4paper, margin=2.cm]{geometry}
\usepackage{amsmath}
\usepackage{nicematrix}
\usepackage{array}
\usepackage{booktabs}
\usepackage{graphicx}
\usepackage{authblk}
\usepackage{setspace}
\usepackage{verbatim}
\usepackage{color,soul}
\usepackage{cite}
\usepackage{multicol}
\usepackage{hyperref}
\hypersetup{
  colorlinks   = true, 
  urlcolor     = blue, 
  linkcolor    = blue, 
  citecolor    = blue  
}
\usepackage{orcidlink}

\DeclareGraphicsExtensions{.png,.pdf}
\title{Reactive molecular dynamics simulations of atenolol first steps degradation by 9CL6 ammonia monooxygenase }

\author[1,2]{Pascal Brault\,\orcidlink{0000-0002-8380-480X}\, \thanks{Corresponding author: pascal.brault@ms4all.eu}}
\author[,1]{Matthieu Wolf\, \orcidlink{0000-0002-6908-9125}}
\author[1]{Yana Gouzi}
\author[3]{Claire Albasi\,  \orcidlink{0000-0001-5101-4611}}
\author[3]{Jeanne Trognon\, \orcidlink{0009-0006-8612-5147}}
\affil[1]{MS4ALL; 1 avenue du champ de Mars, F-45100 Orléans, France}
\affil[2]{GREMI, Université d'Orléans, CNRS; 14 rue d'Issoudun, F-45067 Orléans, France }
\affil[3]{Laboratoire de Génie Chimique, Université de Toulouse, CNRS, INPT, UPS, F-31062 Toulouse, France}

\date{\today}

\begin{document}

\maketitle

\begin{abstract}
The persistence of beta-blockers like atenolol in aquatic environments necessitates efficient remediation strategies, such as enzymatic biodegradation. This study employs reactive molecular dynamics (rMD) simulations to investigate the initial degradation steps of atenolol by the oxidoreductase 9CL6 ammonia monooxygenase. Using the ReaxFF reactive forcefield within a simulated aqueous environment, the degradation processes were modeled at 300 K and 350 K over 6.5 nanoseconds. Results indicate initial degradation rates of 24\% at 300 K and 18\% at 350 K, with each enzyme molecule degrading approximately 20 atenolol molecules before reaching a saturation point. Mass spectrum analysis reveals the primary formation of C8 and C6 fragments at both temperatures. However, at 300 K, additional intermediate fragments (C1, C3, C11, C13) are observed, which are completely bypassed at the higher temperature of 350 K. Furthermore, increasing the temperature to 350 K accelerates the initial degradation phase from 2 ns to 0.5 ns, despite yielding a slightly lower overall degradation rate. Ultimately, these atomistic insights provide a foundational computational methodology for predicting biodegradation rates and pathways of pharmaceutical contaminants at the molecular scale. 
\end{abstract}

\newpage
\section{Introduction}
The widespread occurrence and persistent accumulation of beta-blockers as emerging contaminants in aquatic environments pose severe ecotoxicological threats to marine and freshwater organisms, necessitating urgent, sustainable remediation strategies. A recent review evaluates the microbial biodegradation of beta-blocker contaminants, such as atenolol, metoprolol, and propranolol, in water environments \cite{Ratnasari2025}. Methodologically, the study synthesizes over 80 articles published between 2017 and 2024 from major scientific databases to assess current biodegradation pathways and processes . The degradation of these pharmaceutical compounds, directly or via co-metabolism, is primarily driven by diverse groups of microorganisms, including archaea, bacteria, fungi, and algae, which transform complex drugs into rxpected less harmful metabolites. Specific biodegrading agents highlight fungi like \textit{G. lucidum} and \textit{T. versicolor} \cite{JaenGil2019}, as well as bacterial genera such as \textit{Hydrogenophaga} \cite{Yi2022, Rezaei2022} and \textit{Pseudomonas} \cite{Granatto2021}. The enzymatic mechanisms facilitating this breakdown involve amidohydrolase in bacteria \cite{Yi2022}, alongside lignin peroxidases and manganese-dependent peroxidases in fungal species \cite{JaenGil2019}. The efficiency of enzymatic biodegradation varies significantly based on the organisms and environmental conditions. For instance, anaerobic co-metabolism achieved an 80\% removal rate for atenolol and 60\% for both propranolol and metoprolol \cite{Tang2020}. Fungal treatments successfully degraded metoprolol acid with up to 50\% efficiency over a 10-day incubation period \cite{JaenGil2019}, while specific communities of archaea and bacteria degraded propranolol by up to 88.4\% without the need for co-substrates \cite{Granatto2021}. Despite these promising efficiency rates, the precise roles of many specific microbial enzymes remain underexplored \cite{Ratnasari2025}, and suggests that utilizing formulated microbial consortia rather than single organisms is the most effective way to optimize overall biodegradation \cite{Rezaei2022}.

Focusing specifically on atenolol, its biodegradation can be achieved by enriched nitrifying sludge through co-metabolism involving ammonia-oxidizing bacteria (AOB) and heterotrophic bacteria \cite{Xu2017}. The primary degradation pathway involves the microbial hydrolysis of the amide bond, leading to the formation of atenolol acid \cite{Xu2017}. The degradation of this atenolol acid is often a rate-limiting step, requiring the cleavage of its ether bond \cite{Yi2022}. However, utilizing specific bacterial strains like \textit{Hydrogenophaga} sp. YM1 can overcome this limitation: the addition of co-substrates such as acetate provides extra alpha-ketoglutarate and reducing power to stimulate the alpha-ketoglutarate-dependent dioxygenase enzyme (TfdA), enabling complete mineralization into carbon dioxide via the TCA cycle \cite{Yi2022}. In terms of bioprocesses, sequencing batch reactors (SBR) have demonstrated high efficiency in treating high loads of this drug \cite{Rezaei2022}. Under optimal conditions (an initial concentration of 400 mg/L and a hydraulic retention time of 40 hours), an SBR achieved 91\% removal of atenolol and 87\% removal of the associated chemical oxygen demand (COD), requiring approximately 80 days for biomass acclimation \cite{Rezaei2022}. This method also revealed an alternative biodegradation pathway resulting in simple linear compounds at the end of the treatment cycle \cite{Rezaei2022}. Finally, environmental conditions dictate the generated by-products: while the presence of ammonium favors the formation of additional metabolites (such as P117 and P167 through ether bond cleavage and N-dealkylation) via AOB cometabolism \cite{Xu2017}, the absence of ammonium limits the degradation to classical metabolic pathways (yielding only P267 and P227) \cite{Xu2017}.

Since primary reaction process of biodegradation occurs at the molecular scale, reactive molecular dynamics simulations (rMD),  based on solving classical Newton equations of motion, are a relevant approach for predicting biodegradation rates. This Letter is intented to build a methodology for addressing initial step of enzymatic biodegradation using rMD. The case under study is the biodegradation of atenolol with the oxidoreductase 9CL6 ammonium monooxygenase, from Nitrosomonas europaea ATCC 19718. It is expected to be efficient, at least for iniating the degradation. The next section is dedicated to the rMD method. The third section presents the results and a discussion. The last section will provide the conclusions of the work.

\section{Methods}
Molecular Dynamics (MD) is a simulation method that describes the behavior of an N-body system. It considers each of the $N$ species in the simulation as distinct objects subjected to Newtonian mechanics, i.e., with a defined position and velocity dependent only on the forces exerted on them, according to the following equation:

\begin{equation}
    \label{eq1}
     m_i \frac{d^{2} \vec{r}_i(t)}{d t^{2}} = \vec{f}_i(t), \quad \mbox{with} \quad
     \vec{f}_i(t) =  -\frac{\partial V(\vec{r}_1(t), \vec{r}_2(t), ...,\vec{r}_n(t))}{\partial \vec{r}_i(t)} 
\end{equation}

Where $\vec{r}_{i}(t)$ is the position of atom $i$ at time $t$, $m_{i}$ is its mass, and $\vec{f}_{i}(t)$ are the forces applied to atom $i$ at time $t$. $V = V(\vec{r}_1, \vec{r}_2, ...,\vec{r}_n)$ is the interaction potential among all species. These forces exerted on each atom then determine its position and velocity at the next time step through discrete time integration. By repeating this method as many times as necessary, it becomes possible to determine the behavior of atom  assembly.\\
To correctly simulate the behavior of these systems, the initial configuration must be known, ideally matching experimental conditions as closely as possible. The forces between atoms are derived from the interaction potential $V(\vec{r}_1(t), \vec{r}_2(t), ...,\vec{r}_n(t))$. In the context of reaction dynamics, a reactive forcefiels is required. The reaxFF family of forcefields \cite{Senftle2016} has demonstrated its relevance in various domains, including studying molecule degradation \cite{Brault2021,Richard2025} and biological materials exposure to reactive species \cite{Neyts2014,Yusupov2015,Yusupov2017}. It is both reactive (allows bond breaking and formation based on bond order concept) and includes variable charges (atomic fractional charges calculated and equilibrated every chosen timestep interval) \cite{Senftle2016}.  
ReaxFF was  originally developed to simulate hydrocarbons \cite{duin2001}. This forcefield considers numerous factors when calculating the energy contribution of the interaction between each atom in the system. The system's energy is calculated as follows:
\[
E_{system} = E_{bond} + E_{over} + E_{under} + E_{val} + E_{pen} + E_{tors} + E_{conj} + E_{vdW} + E_{Coulomb}
\]
$E_{bond}$ uses the distance between two atoms to determine the bond order and calculate its energy. The terms $E_{over}$ and $E_{under}$ impose energy penalties when an atom has too many or too few bonds with its neighbors. Additionally, $E_{val}$ and $E_{tors}$ add an energy penalty to the system when the valence and torsion angles deviate from the expected equilibrium value, respectively. $E_{pen}$ is present to penalize certain configurations when two double bonds share the same atom. The term $E_{conj}$ is used to account for conjugated systems, while $E_{vdW}$ represents the van der Waals interaction between different atoms. Finally, $E_{Coulomb}$ represents the Coulomb interactions between atoms due to the partial charges assigned to them by the "Electronegativity Equalization Method" \cite{Janssens1997,Mortier2002}.
In this study, we used the reaxFF "glycine" parameter file \cite{Rahaman2011}, relevant for the sytem under study. 

The simulation box is a cube of 70 x 70 x 70 \AA$^3$, including the 9CL6 enzyme, 100 atenolol and 2000 H$_2$O molecule, for mimicking an aqueous environnement: the density of the system is 0.8 g.cm$^{-3}$.
The simulations are carried out at 300 K (consistent with usual operating temperature) and 350K (for accelerating kinetics). The temperature are maintained using a Langevin thermostat with damping time of 25 fs, \emph{i.e.} 100 times the timestep (dt = 0.25 fs), used for solving the Newton equations of motion. A simulation lasts for 26. 10$^6$ timesteps \emph{i.e.} 6.5 ns.
\begin{figure}[ht]
    \centering
\includegraphics[scale=0.7]{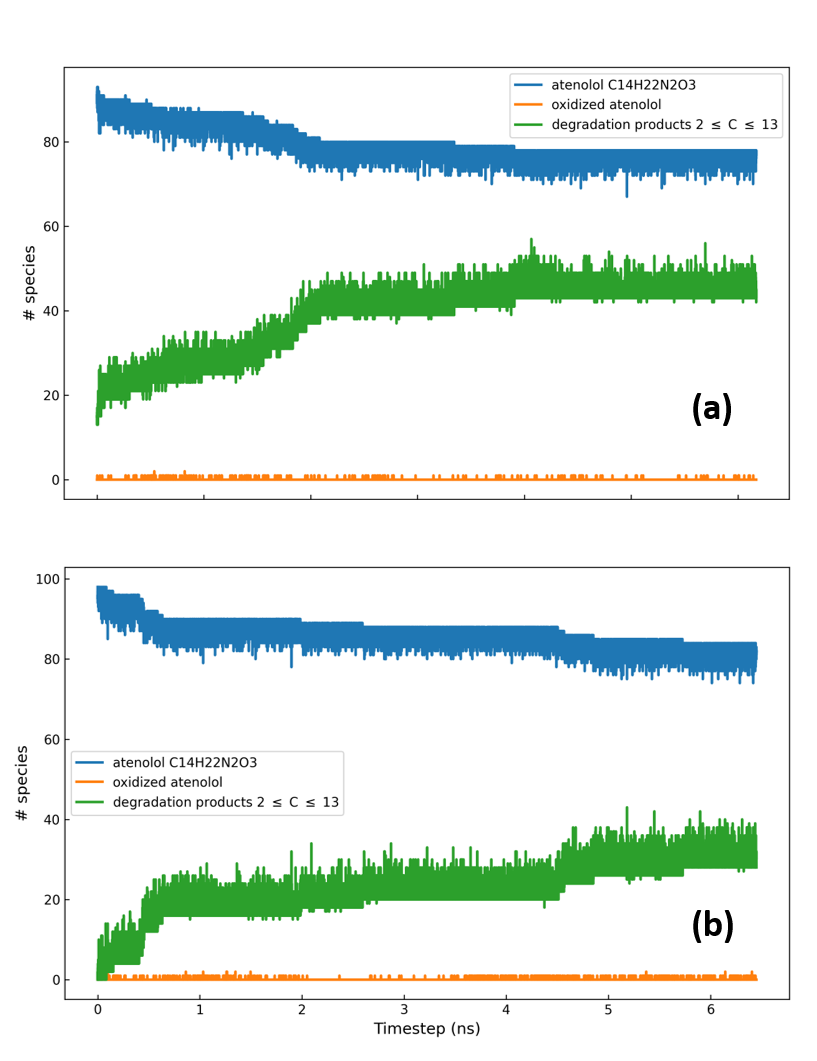}
    \caption{Plot of the atenolol degradation, product production and oxidized atenolol (a) 300K, (b) 350 K}
    \label{fig:1}
\end{figure}
\section{Results and discussion}
\begin{figure}[ht]
    \centering
\includegraphics[scale=0.7]{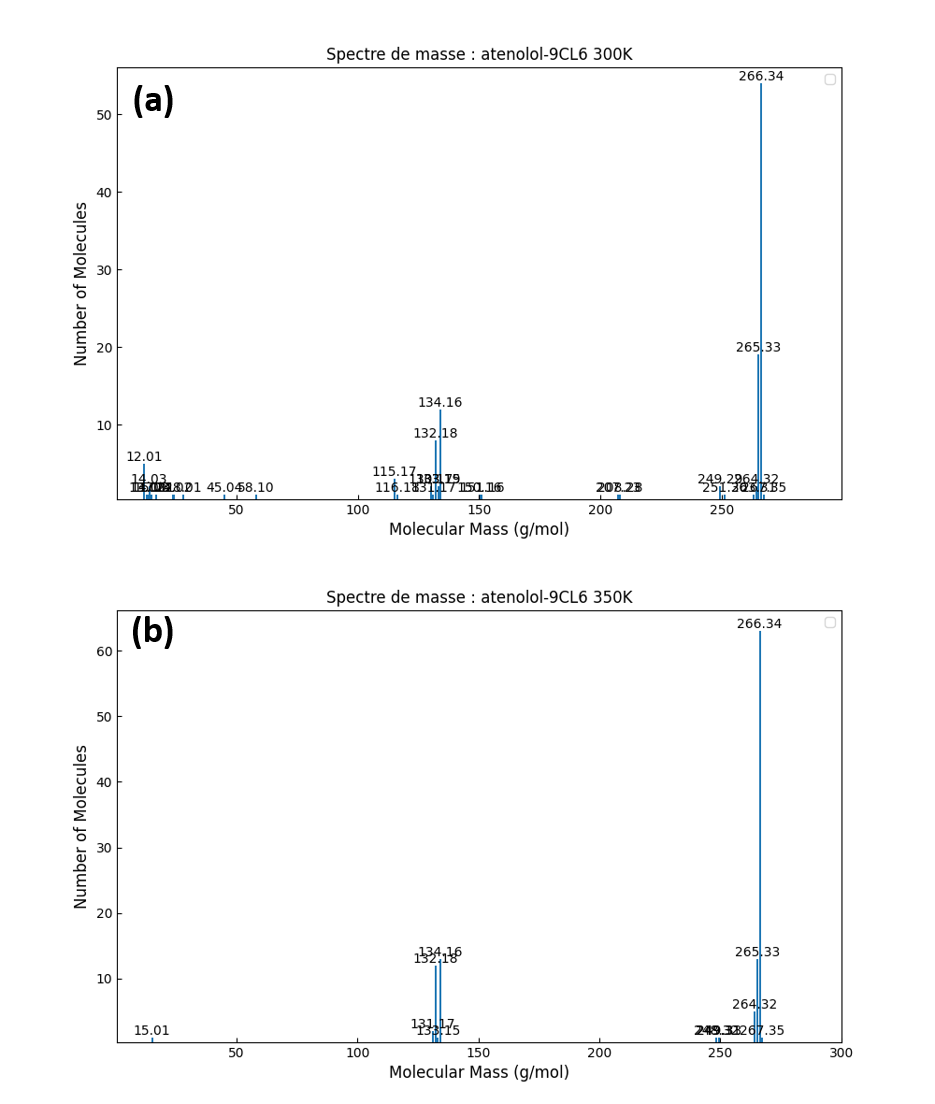}
    \caption{Mass spectra at the end of the simulation (a) 300 K ,  (b) 350 K}
    \label{fig:2}
\end{figure}

Figure \ref{fig:1} displays the degradation of atenolol and the associated product formation at 300 and 350 K. After  26. 10$^6$ timesteps degradation rate reachs 24\% at 300 K and 18\% at 350K, what appears to be a saturation point in biodegradation. These values are close together considering the fluctuation width of the curves. One could guess that increasing the number of atenolol molecules would reduce the fluctuations. We verified it is not the case, since no change in the degradation rate occured. It means that a single enzyme is not able, in the calculation, to degrade more than around 20 atenolol molecules. Increasing biodegradation would require a simulation box with more 9CL6 enzymes. It is noticeable that there is no oxidized atenolol species. Observation of the degradation by 9CL6 enzyme is consistent with experimental findings \cite{Xu2017 ,Yi2022,Rezaei2022}, but without atenolol acid production.
Product formation is detailed by examining the simulated mass spectrum (Figure \ref{fig:2}). Tables \ref{tab:1} and \ref{tab:2} gives the molecule assignement to mass. It should be noticed that products with same carbon, oxygen and nitrogen number can differ by a few H atoms, and should  be considered as the same molecule: \emph{e.g.} molecules C$_{14}$H$_{19-23}$O$_3$N$_2$ are the same molecule, atenolol in Tables \ref{tab:1} and \ref{tab:2}.

Biodegradation at 300K leads to more fragments than at 350 K, without any product at masses greater than atenolol. At 300K, C$_8$ and C$_6$ species are the main products, but also with intermediate C$_1$ and C$_{13}$, C$_3$ and C$_{11}$ products. At 350 K the products are only C$_8$ and C$_6$. Further fragmentation would require more time and/or O$_2$ addition in the simulation box. Increasing the temperature results in by-passing C$_1$ and C$_{13}$, C$_3$ and C$_{11}$ product formation. This is also visible in \ref{fig:1}, where degradation and product formation is reduced from 2 ns at 300 K to 0.5 ns at 350 K. The 9CL6 is thought to have an optimal temperature of approximately 30–35°C (which falls between 300 and 350 K), which also explains the decrease in the degradation rate \cite{Groeneweg1994}. On the other hand, it is not sure that at 350 K, the enzyme’s 3D conformation remains intact, which would explain the decrease in atenolol degradation as well as the number of fragments produced. 

\begin{table}[ht]
\caption{Degradation products at the end of the simulation at 300 K}
\label{tab:1}
\centering
\begin{tabular}{c c c } 
 \hline \hline    
Molecule	& Molecular Mass (g/mol)&	Number of Molecules\\
\hline
C14H23O3N2 &	267.3454 &	1\\
C14H22O3N2&	266.3374	&54\\
C14H21O3N2&	265.3294&	19\\
C14H20O3N&2	264.3214&	2\\
C14H19O3N2&	263.3134& 1\\
C13H19O3N2&	251.3027&	1\\
C13H17O3N2&	249.2867&	2\\
C12H18O2N&	208.2779&	1\\
C11H13O3N&	207.2266&	1\\
C8H9O2N&	151.1631&	1\\
C8H8O2N&	150.1551&	1\\
C8H8ON&	134.1557&	12\\
C6H15O2N&	133.1897&	2\\
C8H7ON&	133.1477&	2\\
C6H14O2N&	132.1817&	8\\
C6H13O2N&	131.1737	&1\\
C6H12O2N&	130.1657&	2\\
C6H14ON&	116.1823&	1\\
C6H13ON	&115.1743&	3\\
C3H8N&	58.1028&	1\\
CH3ON&	45.0408&	1\\
CO&	28.0101&	1\\
C2&	24.0214&1\\
HO&	17.0074	&1\\
CH3&	15.0347&	1\\
CH2	&14.0267&	2\\
N&	14.0067&	1\\
CH&13.0187	&1\\
C&	12.0107&	5\\
\hline
\end{tabular}
\end{table}

\begin{table}[ht]
\caption{Degradation products at the end of the simulation at 350 K}
\label{tab:2}
\centering
\begin{tabular}{c c c } 
 \hline \hline    
Molecule	& Molecular Mass (g/mol)&	Number of Molecules\\
\hline
C14H23O3N2	&267.3454&	1 \\
C14H22O3N2	&266.3374&	63 \\
C14H21O3N2	&265.3294&	13\\
C14H20O3N2	&264.3214&	5\\
C14H21O2N2	&249.33&	1\\
C14H19O3N	&249.3067&	1\\
C14H20O2N2	&248.322&	1\\
C8H8ON	&134.1557&	13\\
C8H7ON	&133.1477&	1\\
C6H14O2N&	132.1817&	12\\
C6H13O2N&	131.1737&	2\\
HN	&15.0147	&1\\
\hline
\end{tabular}
\end{table}

\section{Conclusion}
This study successfully establishes a computational methodology using reactive molecular dynamics (rMD) to model the initial biodegradation steps of the beta-blocker atenolol by the 9CL6 ammonia monooxygenase enzyme. Our simulations in an aqueous environment demonstrate that this enzyme effectively initiates degradation, reaching completion rates of 24\% at 300 K and 18\% at 350 K within a 6.5 ns timeframe. A key finding is the identification of a biodegradation saturation point, where a single 9CL6 enzyme is limited to degrading roughly 20 atenolol molecules. Because increasing the number of atenolol molecules in the simulation did not alter the degradation rate, scaling up the biodegradation process would necessitate higher enzyme concentrations.  Furthermore, mass spectra analyses confirm that the degradation pathways and resulting by-products are highly temperature-dependent. At 300 K, the enzymatic breakdown yields a diverse array of fragments, including primary C8 and C6 species alongside smaller C1, C3, C11, and C13 intermediates. Conversely, elevating the temperature to 350 K accelerates the initial degradation phase—reducing the reaction time from 2 ns to 0.5 ns—but limits the product variety exclusively to C8 and C6 fragments, effectively bypassing intermediate species formation.  Notably, the absence of oxidized atenolol and atenolol acid among the simulated by-products aligns with certain experimental observations but also highlights the limitations of the current simulation scope. To achieve deeper fragmentation and study further metabolic breakdown, the simulation parameters must be expanded. Future studies should consider extending the simulation timeframe or introducing molecular oxygen into the simulation box to evaluate subsequent oxidative degradation steps. In conclusion, this rMD approach proves highly relevant for predicting initial enzymatic degradation rates and uncovering the fundamental molecular mechanisms required for treating pharmaceutical wastewater.
\section*{Acknowledgements}
This work was granted access to the HPC resources of TGCC under the allocation 2025-AD010816938 made by GENCI (Grand Equipement National de Calcul Intensif) .\\
\bibliographystyle{unsrt}
\bibliography{atenolol_rMD} 

\end{document}